\documentclass[aps,prl,twocolumn]{revtex4}

\usepackage[normalem]{ulem}

\usepackage{amsmath}
\usepackage{amsfonts,amssymb,amsthm, bbm,braket}
\usepackage{graphicx}   
\usepackage{subfigure}  
\usepackage{amsbsy} 
\usepackage[bold]{hhtensor} 
\usepackage{natbib}
\usepackage{tipa}
\usepackage{algpseudocode}

\usepackage{multirow}
\usepackage{tensor} 

\newcommand\Tr{\mathrm{Tr}}

\usepackage[breaklinks=true]{hyperref}
\hypersetup{
  colorlinks   = true, 
  urlcolor     = blue, 
  linkcolor    = blue, 
  citecolor   = red 
}

\begin{document}

\title{The best Fisher is upstream: data processing inequalities for quantum metrology}

\author{Christopher Ferrie}
\affiliation{
Center for Quantum Information and Control,
University of New Mexico,
Albuquerque, New Mexico, 87131-0001}

\date{\today}


\begin{abstract}
We apply the classical data processing inequality to quantum metrology to show that manipulating the classical information from a quantum measurement cannot aid in the estimation of parameters encoded in quantum states.  We further derive a quantum data processing inequality to show that coherent manipulation of quantum data also cannot improve the precision in estimation.  In addition, we comment on the assumptions necessary to arrive at these inequalities and how they might be avoided providing insights into enhancement procedures which are not provably wrong. 
\end{abstract}


\maketitle

Parameter estimation is an integral part of physics.  Quantum metrology refers to the study of the ultimate limits in the accuracy of estimates given the structure imposed by quantum theory \cite{caves1981quantum,giovannetti2011advances}. Estimation at or near this limit is important for practical objectives such as improving time and
frequency standards \cite{udem2002optical,hinkley2013atomic} as well as fundamental physics, such as the detection of gravitational waves \cite{aasi2013enhanced}.  

Researchers have found many novel approaches to quantum metrology using, for example, multi-pass interferometers \cite{Higgins2007Entanglement}, machine learning techniques \cite{Hentschel2011Efficient} and computational Bayesian statistics \cite{Granade2012Robust} as well as new bounds \cite{Boixo2007Generalized,Tsang2011Fundamental} in increasingly more general scenarios.  Here we supplement these results with one of a different flavor.  We provide a very general bound on the estimation accuracy in quantum metrology when noise or data processing (either classical or coherent) are present.  Precisely, we give a classical and quantum \emph{data processing inequality} which shows that estimators based on the raw data are optimal.  In other words, processing the data cannot improve quantum metrology.  We conclude by showing how to avoid the inequalities with more exotic procedures which can be classed into three conceptually intuitive categories: (1) processing data in a way dependent on the parameter; (2) circumventing an imposed operational restriction; or (3) modifying the dynamics which impart the parameter.

Consider a statistical model defining a likelihood function $\Pr(x|\theta;C)$.  In words, there is an experimental context $C$ whose outcomes are labeled by the random variable $x$ and $\theta$ is an unknown parameter to be estimated.  For quantum metrology, the goal is estimate a parameter which defines a quantum dynamical process:
\begin{equation}\label{eq:impose}
\rho \mapsto \rho(\theta) := \sum_j K_j(\theta)\rho K_j^\dag(\theta), \; \sum_j K_j^\dag(\theta)K_j(\theta) = \mathbbm 1.
\end{equation}
The statistical model is given by the structure of quantum theory and the Born rule:
\begin{equation}
\Pr(x|\theta;\{E_k\},\rho) = \Tr(\rho(\theta) E_x),
\end{equation}
where the set $\{E_k\}$ forms a quantum measurement which defines the chosen detection strategy.  In broad strokes, the goal of quantum metrology is to find the experiment context $C = (\rho, \{E_k\})$ which allows for the best accuracy in estimating $\theta$.  But how do we measure accuracy?  The standard metric is \emph{mean squared error}:
\begin{equation}\label{risk}
R(\theta, \hat \theta;C) = \mathbb E_{x|\theta;C}[|\theta-\hat\theta(x;C)|^2],
\end{equation}
where $\hat \theta$ is an \emph{estimator}, a function which takes every possible data set to an estimate of $\theta$.  Note we have used the notation $\mathbb E_z[f(z)]$ to mean the expectation of the function $f$ with respect to the distribution of $z$.  The symbol `R' stands for `risk' and Eq.~\eqref{risk} denotes the risk of using the estimator $\hat\theta$ when the true parameter is $\theta$.  

One of the conveniences of using squared error as a measure of loss is that the risk can be lower bounded using the \emph{Cramer-Rao bound} (CRB) \cite{Lehmann1998Theory}:
\begin{equation}
R(\theta, \hat \theta;C) \geq I(\theta;C)^{-1},
\end{equation}
where $I(\theta;C)$ is the \emph{Fisher information}:
\begin{equation}
I(\theta;C) = \mathbb E_{x|\theta;C}\left[ \left(\frac{\partial}{\partial\theta} \log \Pr(x|\theta;C)\right)^2\right].
\end{equation}
The CRB is a fundamental and powerful tool in statistical estimation since it bounds the performance of \emph{every} unbiased estimator.  Although the bound generally depends on the true value of the parameter, for many quantum metrology problems considered so far in the literature the Fisher information has been independent of the unknown parameter.  However, this is not generally true and we must take account of the fact that $\theta$ is unknown and perhaps itself a random variable.

Suppose then that $\theta$ is a random variable with probability density $\Pr(\theta)$.  Then we can remove the dependence of the risk on $\theta$ by taking a second average:
\begin{equation}\label{bayes risk}
r(C) = \mathbb E_\theta[R(\theta,\hat\theta;C)].
\end{equation}
The reason that $r$ does not depend on the estimator $\hat \theta$ is that it is well-known in statistics that the unique estimator which minimizes this quantity is \cite{Lehmann1998Theory}
\begin{equation}\label{bme}
\hat \theta(x;C) = \mathbb E_{\theta|x;C}[\theta].
\end{equation}
Using this, the expression for $r$ can be simplified to 
\begin{align}
r(C)& = \mathbb E_\theta[\mathbb E_{x|\theta;C}[|\theta -\hat\theta(x;C)|^2]],\\
&=\mathbb E_{x;C}[\mathbb E_{\theta|x;C}[|\theta -\hat\theta(x;C)|^2]],\\
&= \mathbb E_{x;C}[{\rm Var}_{\theta|x;C}[\theta]].\label{postvar}
\end{align}
where Var denotes the variance.  Note that $\Pr(\theta|x;C)$ is the \emph{posterior} distribution using Bayes rule:
\begin{equation}
\Pr(\theta|x;C) = \frac{\Pr(x|\theta;C)\Pr(\theta)}{\Pr(x;C)}.
\end{equation}
For this reason, $r(C)$ is called the \emph{Bayes risk}, which we have shown in Eq.~\eqref{postvar} is the expected posterior variance, and $\hat \theta(x;C)$ is called the \emph{Bayes estimator}.  The Cramer-Rao bound is also generalized to the \emph{Bayesian Cramer-Rao bound} (BCRB) \cite{Gill1995Applications}:
\begin{equation}
r(C)\geq J(C)^{-1},
\end{equation}
where $J$ is the \emph{Bayesian information}:
\begin{equation}
J(C) = \mathbb E_{\theta}[I(\theta;C)].
\end{equation}
Note that everything stated above generalizes in the expected way when $\theta \in \mathbb R^d$ is a vector of unknown parameters.

As stated above, quantum metrology seeks to find the experimental context which minimizes the \emph{risk}.  Or, since the bounds stated above are generally achievable (at least asymptotically), we seek to maximize the \emph{information}.  For example, the quantity
\begin{equation}\label{qfi}
I_{\rm Q}(\theta) = \max_{C} I(\theta;C)
\end{equation}
we call the \emph{quantum Fisher information}.  We can also define the quantity
\begin{equation}\label{qbi}
J_{\rm Q} = \max_{C} J(C),
\end{equation}
which we analogously call the \emph{quantum Bayesian information}.  Using these we have two \emph{quantum} Cramer-Rao bounds:
\begin{align}
R(\theta,\hat\theta)&\geq I_{\rm Q}(\theta),\\
r &\geq J_{\rm Q}.
\end{align}
These inequalities place the ultimate limit (called the \emph{Heisenberg limit}) on the estimation accuracy of the unknown parameter $\theta$.  Operationally, the location of the maxima in equations Eqs.~\eqref{qfi} and \eqref{qbi} specify the physical \emph{experiment} which must be performed to achieve this ultimate limit.

For a fixed state $\rho$, the optimization over the measurement alone in Eq.~\eqref{qfi} was introduced by Braunstein and Caves \cite{Braunstein1994Statistical} and shown to be equivalent to the original definition of the quantum Fisher information given by Helstrom \cite{Helstrom1976Quantum}:
\begin{equation}\label{sld fisher}
\tensor[_{\rm sld}]{I}{_{\rm Q}}(\rho(\theta)) = \Tr(\rho(\theta) L(\theta)^2), 
\end{equation}
where the operator $L$, the \emph{symmetric logarithmic derivative} (SLD),  is implicitly defined via
\begin{equation}
\frac{\partial}{\partial \theta} \rho(\theta) = \frac12(\rho(\theta)L(\theta)+L(\theta)\rho(\theta)).
\end{equation}
To distinguish it from the more general definition in Eq.~\eqref{qfi}, we call the definition in Eq.~\eqref{sld fisher} the SLD Fisher information.  As noted, the crucial difference is that the SLD Fisher information depends on $\rho$---that is, it is assumed that the choice of initial state is fixed.  For this reason, we prefer Eq.~\eqref{qfi} (or Eq.\eqref{qbi} in the Bayesian context) since it makes clear that $\theta$ is unknown and $C$ is an experimental context, the design of the \emph{full} experiment.  This also allows us to easily restrict $C$ when physical or practical constraints are present (such as local measurements or Gaussian states).  It also makes clear that the state is part of the design, which in the general case must simultaneously be optimized \cite{hayashi2011comparison}.  On the other hand, in many cases the optimization of the measurement and preparation context can be performed separately \cite{Lang2013Optimal}, thus making the SLD Fisher information a powerful calculation tool in such cases.

Another important reason to prefer the definition of the quantum Fisher information in Eq.~\eqref{qfi} as opposed to the symmetric logarithmic derivative version in Eq.\eqref{sld fisher} is that the latter is not general achievable for more than a single parameter $\theta$.  In other words, to achieve the Fisher information $\tensor[_{\rm sld}]{I}{_{\rm Q}}$ may require incompatible measurements \cite{Helstrom1976Quantum}.  The definition in Eq.\eqref{qfi} explicitly restricts the information to that achievable by valid quantum mechanical measurements.

Finally, we note that Eqs.\eqref{qfi} and \eqref{qbi} are \emph{operational}---they tell us exactly what experimental context maximizes the information content of the measurement.  With these operational definitions of information we give a more enlightening and operational definition of ``Heisenberg limit'', which is necessarily problem dependent: given a specification of the problem, \emph{Heisenberg limited metrology} is a realization of the experimental designs required to achieve the maximum information in either Eq.~\eqref{qfi} or ~\eqref{qbi}.  This operational definition alleviates the need to resolve the recent confusion of the term \cite{zwierz2010general}; the Heisenburg limit cannot be beaten because it is the limit, by definition.  We can also consider restricting the allowed context $C_r\subset C$ such that the optimum cannot be achieved.  For example, we could impose a restriction to laser sources and photon number constraints \cite{Lang2013Optimal}.  

Another relevant restriction $C_r$ is to that of product state inputs and outputs.  In this case, the maximization of $I(\theta;C)$ or $J(C)$ over $C_r$ is typically called the ``standard quantum limit'' \cite{giovannetti2004quantum}.  In the special case of a restriction to independent trials, it is called the ``shot noise limit''.  If we call such restrictions ``classical'', we implicitly define a  \emph{quantum resource}: those experimental contexts in $C\backslash C_r$ whose information is larger than that maximized over $C_r$.

Having specified the problem, we will now apply the so-called \emph{data processing inequality} to the quantum metrology to show that post-processing of the data can never improve the estimation accuracy.  First, a definition: $\theta\to x \to y$ is called a \emph{Markov chain} if $\Pr(y,x,\theta) = \Pr(y|x)\Pr(x|\theta)\Pr(\theta)$.  Note that if $y$ is some deterministic function (a \emph{statistic}) of $x$, that is $y = f(x)$, then  $\theta\to x \to f(x)$ is trivially a Markov chain.  Why is this relevant to estimation?  The chain $\theta\to x \to y$ can be thought of as an estimation procedure where $\theta$ generates the raw data $x$ via the statistical model $\Pr(x|\theta)$ and then that data is \emph{post-processed} (in general probabilistically) to arrive at $y$.  The information flowing through the chain can be used to estimate $\theta$.  Next, we show that the second step, post-processing, cannot improve the estimation accuracy.

The first data processing inequality applies to the Fisher information and is \cite{Rioul2011Information}
\begin{equation}\label{dpi1}
I_y(\theta) \leq I_x(\theta),
\end{equation}
with equality if and only if $\theta\to y\to x$ is also a Markov chain (which is equivalent in the case $y=f(x)$ to $f$ being a \emph{sufficient statistic}).  This inequality implies the analogous Bayesian information variant:
\begin{equation}\label{dpi2}
J_y\leq J_x.
\end{equation}
Both inequalities state that the Fisher (respectively, Bayesian) information calculated using the distribution of processed data $\Pr(y|\theta)$ is less than that computed using the original distribution of raw data $\Pr(x|\theta)$.  Then the Cramer-Rao bounds state the mean squared error of an unbiased estimator of $\theta$ is worse when post-processing.  

Let us apply this to the quantum metrology setting where the conclusion should be unsurprising.  Indeed, it is quite simple to include an addition experimental context $C$ in the classical description above.  Let us start with the Fisher information version first.  The data processing inequality in Eq.~\eqref{dpi1} remains unchanged when adding an additional context:
\begin{equation}
I_y(\theta;C) \leq I_x(\theta;C).
\end{equation}
Since this holds for all $C$, it holds where each side individually obtains its maximum.  That is
\begin{equation}
\max_C I_y(\theta;C) \leq \max_C I_x(\theta;C).
\end{equation}
These are the quantum Fisher informations when using either the raw data $x$ or post-processed data $y$:
\begin{equation}
I_{y,Q}(\theta) \leq I_{x,Q}(\theta).
\end{equation}
Then, the quantum Cramer-Rao bound implies that conditioning on post-processed data cannot improve the estimation of $\theta$.  The same argument applies to Eq~\eqref{dpi2}.  If we add the context $C$ and maximize, we find
\begin{equation}
J_{y,Q} \leq J_{x,Q}.
\end{equation}
The Bayesian Cramer-Rao bound then implies that the Bayes risk of using post-processed data is higher.

The above results apply to the case where ``data processing'' refers to classical computation of classical data.  Perhaps it might be the case that coherent data processing---quantum computation of quantum data---might aid in the estimation of the parameters $\theta$.  In this case, rather than the classical process $\theta\to x\to y$, we have the quantum process $\theta\to \rho(\theta) \to \mathcal E(\rho(\theta))$ , where $\mathcal E$ is a quantum operation (completely-positive, trace preserving map).  Next, we prove a \emph{quantum data processing inequality} which analogously shows that coherent manipulation of data also cannot aid quantum metrology.

The result is as follows.  If $\mathcal E$ is a quantum operation, then 
\begin{equation}\label{qdpi}
I_{\mathcal E,Q}(\theta)\leq I_Q(\theta).
\end{equation}
The proof is remarkably simple.  First note that 
\begin{align}
\Pr(x|\theta;\{E_k\},\mathcal E(\rho)) &= \Tr(\mathcal E(\rho(\theta)) E_x),\\
& = \Tr(\rho(\theta) \mathcal E^\dag(E_x)),\\
&=\Pr(x|\theta;\{\mathcal E^\dag(E_k)\},\rho),
\end{align}
where $\mathcal E^\dag$ is the \emph{dual channel}---a \emph{Heisenburg picture} for quantum channels.  Explicitly, if the map $\mathcal E$ has the Kraus decomposition
\begin{equation}
\mathcal E(\cdot) = \sum_j K_j\cdot K_j^\dag, 
\end{equation}
then
\begin{equation}
\mathcal E^\dag(\cdot) = \sum_j K_j^\dag\cdot K_j.
\end{equation}
In words, the act of subjecting $\rho(\theta)$ to an addition quantum channel is equivalent to subjecting the measurement to the dual channel.  

Now, since it is the measurement to be optimized, either the range of $\mathcal E^\dag$ contains the optimal measurement, or it does not.  The channel $\mathcal E^\dag$ serves only to restrict the possible measurements.  That is,
\begin{equation}
\max_{\rho,\{\mathcal E^\dag(E_k)\}} I(\theta; \rho,\{\mathcal E^\dag(E_k)\})\leq \max_{\rho,\{E_k\}} I(\theta; \rho,\{E_k\}).
\end{equation}
Thus, by definition,
\begin{equation}
I_{\mathcal E,Q}(\theta)\leq I_Q(\theta).
\end{equation}
This is the quantum data processing inequality and it states that no coherent manipulation of the data allowed by quantum theory improves the estimation accuracy of $\theta$.

Some comments are in order.  First, we note the that the temporal order of the data processing is irrelevant.  The quantum process $\theta\to \rho(\theta) \to \mathcal E(\rho(\theta))$ has the channel $\mathcal E$ act after the parameter has been imparted.  However, the conclusion remains if the process is $\rho \to \mathcal E (\rho) \to \mathcal E(\rho)(\theta)$.  That is, Eq. ~\eqref{qdpi} holds if $\mathcal E$ refers to ``pre-processing'' or ``encoding''.

Secondly, we comment on the the terminology ``data processing inequality''.  This term will is more popularly used in the context of information theory, where it applies to the \emph{mutual information} between either $y$ and $\theta$ or $x$ and $\theta$ in the Markov chain $\theta\to x\to y$.  If $\mathcal I(a;b)$ denotes the mutual information (a measure of correlations) between $a$ and $b$ then the more commonly used \emph{data processing inequality} is $\mathcal I(\theta;y)\geq \mathcal I(\theta;x)$ (see, for example, \cite{Cover2012Elements}).  In words, it says the same thing as the inequality we have used here (proven in \cite{Rioul2011Information}): manipulating the data, cannot increase the amount of information one has about $\theta$.  This information theoretic data processing inequality is not directly applicable to estimation but is a fundamental result in information theory.  As one might expect, then, it has been generalized to the quantum mechanical setting \cite{Schumacher1996Quantum}.

The next thing to mention is noise.  Note that, in the classical setting the only assumption was that $\theta \to x\to y$ was a Markov chain.  It need not be the case that $y$ is some deterministic function of $x$.  So, the channel $x \to y$ could also represent classical technical noise on the detector.  So long as the noise is statistically independent of the unknown parameter $\theta$ given $x$, the data processing inequality applies.  Thus, noise assisted metrology cannot be realized.  Similarly, in the quantum channel setting, $\mathcal E$ could represent a decoherence mechanism rather than a purposefully built quantum circuit.  The conclusion remains; decoherence cannot improve estimation accuracy.
  
The final comment is on ``outs''.  How do we avoid this conclusion?  The three most natural possibilities are as follows: (1) have $\mathcal E$ depend on $\theta$; (2) arrange for $\mathcal E$ to  circumvent an additional imposed restriction on the allowed context $C$; or (3) modify the dynamics in Eq.~\eqref{eq:impose} which impose the parameter.  A simple example should help illustrate these approaches.  Suppose we have a qubit and an unknown rotation $\theta$ about the $z$--axis in the Bloch sphere.  The optimal input and output states for this problem are $|+\rangle$ which can be shown to give Fisher information $I = 4$.  Now, the data processing inequality states that no channel $\mathcal E$ applied to $e^{-i\theta\sigma_z}|+\rangle$ can increase this Fisher information.

\begin{figure}[h]\centering
  \includegraphics[width=1\columnwidth]{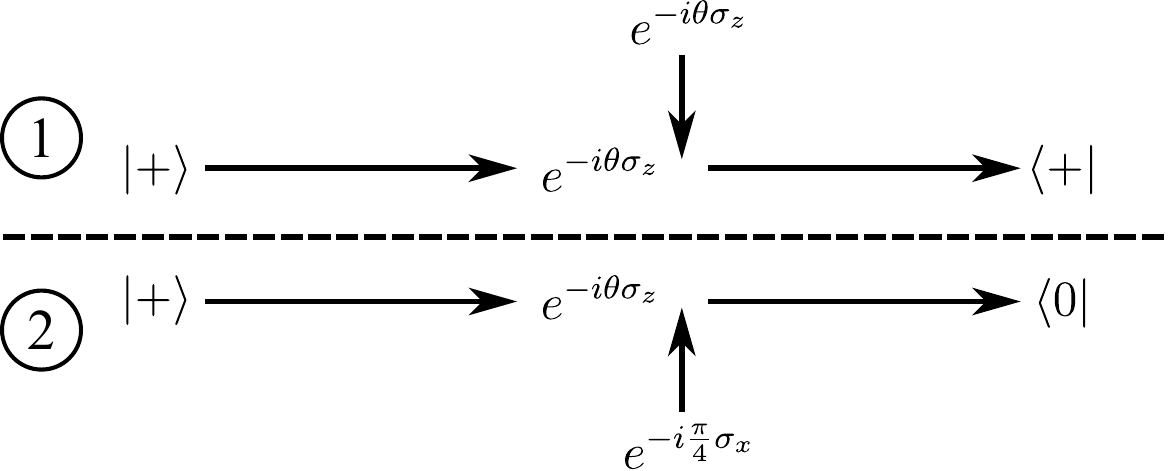}
  \caption{\label{fig:eg} Examples of the two ways to circumvent the data processing inequalities.  Either the additional channel depends on the unknown parameter or it frees us from an operationally imposed restriction. }
\end{figure}

However, situation (1) avoids this conclusion by having the channel depend on $\theta$.  For example, by letting $\mathcal E_\theta = e^{-\theta \sigma_z}$, the rotation is applied again and the Fisher information becomes $I = 16$.  This is depicted in Fig.~\ref{fig:eg} and conceptually equivalent to the multi--pass interferometer of Ref.~\cite{Higgins2007Entanglement}. 

The alternative in situation (2) requires we have a restriction on the experimental context.  Such a restriction (only Gaussian states or only local measurements, for example) could come from physical or operational constraints.  In our toy example, suppose the initial state is restricted to be $|+\rangle$ and the measurement is restricted to be in the $\sigma_z$ basis.  Since the rotation is about $z$, no information can be learned and the Fisher information is indeed $I=0$.  However, suppose we apply the channel $\mathcal E = e^{-i\frac\pi 4\sigma_x}$.  Although this channel does \emph{not} depend on $\theta$, it does avoid the restriction by rotating the state to the measurement plane (or, equivalent in the dual picture, it rotates the measurement to the state plane) and the Fisher information increases to $I=4$.

In situation (3), we are imagining something conceptually different \footnote{The distinction between (1) and (3) is subtle: (1) can be thought of as a subset of (3) but not always vice versa.}.  Here we changing the problem itself.  For example, if it is known that dynamics in Eq.~\eqref{eq:impose} contain certain decoherence terms, learning proceeds at a suboptimal rate \cite{dorner2009optimal}.  A simple, if glib, way to avoid this problem is to remove the decoherence (easy, right?).  A more sophisticated approach, recently rediscovered, is to dynamically correct errors, interleaving the imposition of the parameter with recovery operations \cite{preskill2000quantum, kessler2013quantum}.

Quantum metrology can be thought of as a purely statistical problem.  Often, thinking of quantum mechanical problems classically leads to paradoxes or, in the very least, is just cumbersome---which is why concepts like the SLD quantum Fisher information exist.  However, if we are careful to avoid the usual pitfalls, rephrasing quantum metrology in classical language allows us to leverage known classical results.  In particular, we have applied the data processing inequalities to show that post-processing raw data cannot lead to more precise estimates of parameters.  The classical picture then allows for a simple generalization, which we have called the quantum data processing inequality, showing that coherent data processing suffers the same restriction.  Finally, the classical representation of these results displays more transparently the assumptions necessary to provide this curtailment thus readily allowing us to provide operationally meaningful statements of how to avoid the inequalities.  We hope these considerations shed light on the myriad of definitions of  ``standard quantum limit'', ``Heisenberg limit'' and so on, and perhaps make conceptually clear why and when one can improve on standard estimation procedures.

\begin{acknowledgements}
The author thanks Carl Caves, Josh Combes, Zhang Jiang and Chris Granade (gr\textipa{@}'ne\textipa{I}d) for helpful discussions.  This work was supported in part by  National Science Foundation Grant Nos.~PHY-1212445 and~PHY-1314763 and by Office of Naval Research Grant No.~N00014-11-1-0082 and by the Canadian Government through the NSERC PDF program.
\end{acknowledgements}

\end{document}